# Searching for Superconductivity in Z=10 Family of Materials

O. P. Isikaku-Ironkwe[1, 2]

[1]The Center for Superconductivity Technologies (TCST)
Department of Physics,
Michael Okpara University of Agriculture, Umudike (MOUAU),
Umuahia, Abia State, Nigeria
and
[2]RTS Technologies, San Diego, CA 92122

## Abstract

Binary and ternary materials with average atomic number (Z) of 10 have average valence electrons of 2.67. The material specific characterization dataset (MSCD) scheme, superconductivity symmetry search rules and the Tc estimation formula, earlier developed are used in this search for binary and ternary superconductors with low Z like $MgB_2$. Here we identify 17 potential superconducting Z=10 materials and estimate their Tcs prior to experimental investigations.

## Introduction

$MgB_2$, a low atomic number (Z =7.33) and low valence electron count (Ne = 2.67) material which proved to be superconducting [1] with Tc = 39K has become a model in the search for similar superconductors with high Tc. However such searches [2 – 14] were based only on structural and electronic similarities with $MgB_2$ and have not yielded any Tcs near 39K. Searching in materials with the same Z and Ne as $MgB_2$ has led to many new predictions [15 - 21] of superconductivity awaiting experimental verification. Searches with Z = 4.67 and Ne = 2.67 yielded LiBC-like materials which are not superconducting [6 - 10]. Searches with Z = 12.67 and Ne =2.67 yielded CaBeSi with Tc less than 2K [4] and NaAlSi [14] with Tc =7K. Except for $CaB_2$ [5], the material series with Z = 10 and Ne = 2.67 has not been well searched for superconductivity. In this paper, we initiate the computational search using a novel non-DFT method [16] that can estimate maximum Tc.

## The Search Framework

Computational search for novel superconductivity has long been recommended [23] as a fast approach to explore the millions of possibilities inherent in the Periodic Table. We have developed one such material specific approach that involves first estimating the presence of superconductivity from the chemical formula and next estimating the transition temperature [16] from the chemical formula data. By studying the material specific correlations of superconductivity with Periodic Table properties such as electronegativity, valence electron count, atomic number and formula weight, we have derived criteria and formula [16] that help us estimate superconductivity and Tc of multi-element materials without recourse to McMillan's Tc formula [24] or Density Functional Theory of superconductivity[25, 26]. In addition, some material specific symmetry rules [16] simplify the process of searching for new materials and superconductors.

## Superconductivity Estimation

In an earlier paper [16] we showed that the presence of superconductivity can be estimated in a material from knowledge of the average valence electron count, Ne, and the average atomic number, Z, of the material by the empirical expression:

$$0.75 < Ne/\sqrt{Z} < 1.02 \qquad (1)$$

The higher the lower bound, the more certain we are of superconductivity at ambient pressure. Materials with Ne/$\sqrt{Z}$ much less than 0.75 may be semiconducting or require high pressure to show superconductivity if the electronegativity is also low.

## Search for Z =10 materials

The family of ternary and 3-atom binaries with Z = 10 are easy to search for superconductivity. The strategy is to create combinations of two or three elements whose average Z is 10 and then to compute the averages for electronegativity and valence electrons.

For a compound with formula $A_pB_qC_rD_sE_t$, where p, q, r, s, t are respectively the numbers of atoms of elements A, B, C, D and E, the averages of electronegativity $\mathcal{X}$, valence electron count, Ne and atomic number, Z, can be expressed as:

$$\mathcal{X} = \frac{p\mathcal{X}_A + q\mathcal{X}_B + r\mathcal{X}_C + s\mathcal{X}_D + t\mathcal{X}_E}{p+q+r+s+t} \tag{2}$$

$$Ne = \frac{pNe_A + qNe_B + rNe_C + sNe_D + tNe_E}{p+q+r+s+t} \tag{3}$$

$$Z = \frac{pZ_A + qZ_B + rZ_C + sZ_D + tZ_E}{p+q+r+s+t} \tag{4}$$

We express the formula weight, Fw, for the compound $A_pB_qC_rD_sE_t$ as:

$$Fw = pFw_A + qFw_B + rFw_C + sFw_D + tFw_E \tag{5}$$

Data from these formulae provide the building blocks of the material specific characterization dataset (MSCD). The transition temperature Tc as derived in [16] is given by:

$$T_c = \mathcal{X}\frac{Ne}{\sqrt{Z}}K_o \tag{6}$$

## MSCD

The material specific characterization dataset (MSCD) scheme [16] provides quantitative information on a material in terms of electronegativity, valence electron count, atomic number and formula weight. The MSCD of a material can be used to fine-tune the material for higher Tc or for comparative analysis with other materials. Details of computation of MSCD using equations (2) to (5) above are given in [16]. The MSCD of 17 materials with Z =10 and Ne =2.67 are shown in Table 1.

## Symmetry Search Rules

A symmetry search rule [16] applicable here is that from the MSCDs, we find that all members of the Z = 10 family also have the same valence electron count, Ne. Thus with Ne and Z the same, their Tcs will be proportional to their electronegativities, $\mathcal{X}$. Using $MgB_2$ as a model whose Tc =39K and for which Ko in equation (1) gives 22.85, we computed Ko in terms of Fw/Z and found that Ko = N(Fw/Z), where N = 3.65. The ratio value of $Ne/\sqrt{Z}$ =0.8433 for members of the Z=10 family of materials strongly suggest that they are potential superconductors. The Tc can be estimated using equation (6) and the minimum and maximum N values, empirically estimated for the Z=10 family.

## Discussion

We have explored binary and ternary compounds with Z =10 and Ne =2.67. We find that the 17 explored members of this group have the same valence electron count and formula weights. Like the Z =4.667 and Z =7.333 families, they have Ne = 2.6667, suggesting that the choice of Z limits and restricts the Ne value you can have. The ratio: Ne/SqrtZ gives us 0.8433, suggesting that these families of materials will be superconducting if their electronegativities are high enough. Some of the materials may not form under ambient conditions and may require high pressure to form and be quenched. $CaB_2$ is particularly interesting since it has been predicted [27] to have a Tc between 45 and 50K. Our estimate is that $CaB_2$ cannot have a Tc higher $MgB_2$. Rather we estimate a maximum Tc of 21.7K, if formed. There are yet many families of binary and ternary materials with different values of Z>10. We have shown in this paper how to computational explore a family for superconductivity and estimate the superconductivity Tc range in the family using equation (6). The Z =10 family thus present a treasure box for exploring superconductivity with varying electronegativity and fixed Ne and Z. The results here also pioneer a new route to superconductivity by design rather than by serendipity.

## Conclusion

Many potential superconductors exist in the family of materials with Z = 10, Ne = 2.67 spectrum of binary and ternary compounds, waiting to be experimentally confirmed. We presented 17 of them in this paper and estimated their Tcs to be between 8K and 24K.

## Acknowledgements

The author acknowledges useful discussions with M. Brian Maple and J. Hirsch at UC San Diego, J.R. O'Brien at Quantum Design and M. J. Schaffer , formerly at General Atomics, at various times in the course of this research. M.J. Schaffer also sponsored much of this work.

### Table

| Material | | $\mathcal{X}$ | Ne | Z | Ne/$\sqrt{Z}$ | Fw | Fw/Z | $N_{min}$, $N_{max}$ | Tc(K) Min | Tc(K)Max |
|---|---|---|---|---|---|---|---|---|---|---|
| 1 | $MgB_2$ | 1.733 | 2.667 | 7.3333 | 0.9847 | 45.93 | 6.263 | 3.65 | ------- | 39 |
| 2 | MgBeSi | 1.833 | 2.667 | 10.0 | 0.8433 | 61.41 | 6.141 | 1.1, 2.5 | 10.4 | 23.7 |
| 3 | $Na_2O$ | 1.767 | 2.667 | 10.0 | 0.8433 | 61.98 | 6.198 | 1.1, 2.5 | 10.2 | 23.1 |
| 4 | KBeN | 1.767 | 2.667 | 10.0 | 0.8433 | 62.12 | 6.212 | 1.1, 2.5 | 10.2 | 23.1 |
| 5 | NaMgN | 1.7 | 2.667 | 10.0 | 0.8433 | 61.31 | 6.131 | 1.1, 2.5 | 9.7 | 22 |
| 6 | LiCaN | 1.667 | 2.667 | 10.0 | 0.8433 | 61.03 | 6.103 | 1.1, 2.5 | 9.4 | 21.5 |
| 7 | $CaB_2$ | 1.667 | 2.667 | 10.0 | 0.8433 | 61.70 | 6.170 | 1.1, 2.5 | 9.5 | 21.7 |
| 8 | CaBeC | 1.667 | 2.667 | 10.0 | 0.8433 | 61.10 | 6.110 | 1.1, 2.5 | 9.4 | 21.6 |
| 9 | $Mg_2C$ | 1.633 | 2.667 | 10.0 | 0.8433 | 60.63 | 6.063 | 1.1, 2.5 | 9.2 | 20.9 |
| 10 | NaAlC | 1.633 | 2.667 | 10.0 | 0.8433 | 61.98 | 6.198 | 1.1, 2.5 | 9.4 | 21.3 |
| 11 | BeBSc | 1.6 | 2.667 | 10.0 | 0.8433 | 64.78 | 6.478 | 1.1, 2.5 | 9.6 | 21.9 |
| 12 | NaBSi | 1.567 | 2.667 | 10.0 | 0.8433 | 61.89 | 6.189 | 1.1, 2.5 | 9.0 | 20.5 |
| 13 | $BeAl_2$ | 1.5 | 2.667 | 10.0 | 0.8433 | 62.97 | 6.297 | 1.1, 2.5 | 8.8 | 19.9 |
| 14 | $Be_2Ti$ | 1.5 | 2.667 | 10.0 | 0.8433 | 65.90 | 6.590 | 1.1, 2.5 | 9.2 | 20.8 |
| 15 | NaBeP | 1.5 | 2.667 | 10.0 | 0.8433 | 62.97 | 6.297 | 1.1, 2.5 | 8.8 | 19.9 |
| 16 | LiNaS | 1.467 | 2.667 | 10.0 | 0.8433 | 62.00 | 6.200 | 1.1, 2.5 | 8.4 | 19.2 |
| 17 | LiMgP | 1.433 | 2.667 | 10.0 | 0.8433 | 62.22 | 6.222 | 1.1, 2.5 | 8.3 | 18.8 |
| 18 | LiAlSi | 1.433 | 2.667 | 10.0 | 0.8433 | 62.01 | 6.201 | 1.1, 2.5 | 8.3 | 18.8 |

Table 1: MSCDs of 17, Z = 10 materials and $MgB_2$. The Ko = N(Fw/Z). For MgB2, N = 3.65. For the Z = 10 materials, we have estimated $N_{min}$ = 1.1 and $N_{max}$ =2.5. From these values we can compute $T_c = \mathcal{X} \frac{Ne}{\sqrt{Z}} K_o$. The constant Ne value for this family and the approximately equal Fw for the family is yet to be explained. Tc min ranges between 8.3k and 10.4K, while Tcmax extends from 18.8K to23.7K. These results need experimental verification.

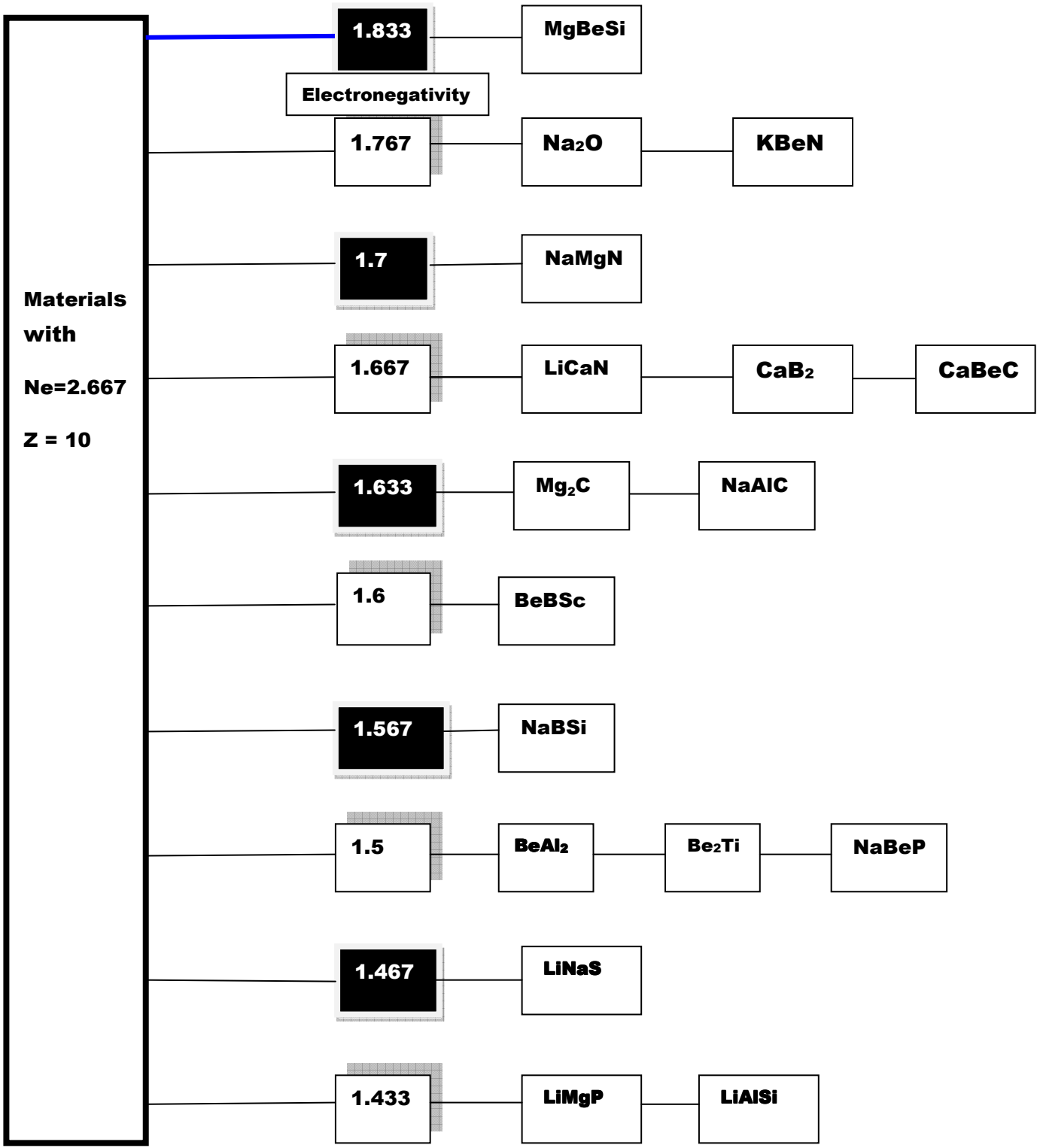

Figure 1: Spectrum of possible Z = 10, Ne =2.667 materials classified under their electronegativities. Since Ne/SqrtZ =0.8433 for this family of materials, we predict that many of them will be superconductors if formed as stable materials. The Tcs can be estimated from $T_c = \mathcal{X} \frac{Ne}{\sqrt{Z}} K_o$ where Ko =N(Fw/Z), Fw represents formula weight[]. N varies between 1.1 to maximum of 2.5 for binary and ternary compounds in this family of Z =10[16].